\documentclass[%
 reprint,
 superscriptaddress,
 amsmath,amssymb,
 aps,
]{revtex4-1}

\usepackage{graphicx}
 \usepackage{dcolumn}
\usepackage{bm}
\usepackage{xcolor}
\usepackage{tabularx}
\usepackage[normalem]{ulem}

\usepackage{hyperref}

\begin{document}

\begin{flushright}
\footnotesize{TTP20-045 P3H-20-083} 
\end{flushright}

\title{Supernova Constraints on Dark Flavored Sectors}

\author{Jorge Martin Camalich}
\affiliation{%
Instituto de Astrof\'isica de Canarias, C/ V\'ia L\'actea, s/n
E38205 - La Laguna (Tenerife), Spain
}%
\affiliation{%
Universidad de La Laguna, Departamento de Astrof\'isica, La Laguna, Tenerife, Spain
}%
\author{Jorge Terol-Calvo}
\affiliation{%
Instituto de Astrof\'isica de Canarias, C/ V\'ia L\'actea, s/n
E38205 - La Laguna (Tenerife), Spain
}%
\affiliation{%
Universidad de La Laguna, Departamento de Astrof\'isica, La Laguna, Tenerife, Spain
}%
\affiliation{%
Instituto de F\'{\i}sica Corpuscular (CSIC-Universitat de Val\`{e}ncia), C/ Catedr\'atico Jos\'e Beltr\'an 2, E-46980 Paterna (Val\`{e}ncia), Spain
}

\author{Laura Tolos}
\affiliation{%
Institute of Space Sciences (ICE, CSIC), Campus UAB,  Carrer de Can Magrans, 08193 Barcelona, Spain
}%
\affiliation{%
Institut d'Estudis Espacials de Catalunya (IEEC), 08034 Barcelona, Spain}
\affiliation{%
Frankfurt Institute for Advanced Studies, Ruth-Moufang-Str. 1, 60438 Frankfurt am Main, Germany
}%

\author{Robert Ziegler}
\affiliation{Institut  f\"ur  Theoretische  Teilchenphysik,  Karlsruhe  Institute  of  Technology,  Karlsruhe,  Germany}%

\begin{abstract}
\noindent Proto-neutron stars forming a few seconds after core-collapse supernovae are hot and dense environments where hyperons can be efficiently produced by weak processes. By making use of various state-of-the-art supernova simulations combined with the proper extensions of the equations of state including $\Lambda$ hyperons,
we calculate the cooling of the star induced by the emission of dark particles $X^0$ through the decay $\Lambda\to n X^0$. 
Comparing this novel energy-loss process to the neutrino cooling of SN 1987A allows us to set a stringent upper limit on the branching fraction, BR$(\Lambda\to n X^0)\leq 8\times10^{-9}$, that we apply to massless dark photons and axions with flavor-violating couplings to quarks. We find that the new supernova bound can be orders of magnitude stronger than other limits in dark-sector models.   
\end{abstract}

\pacs{Valid PACS appear here}

\maketitle

\textbf{\textit{Introduction:}} Clarifying the fundamental nature of dark matter remains one of the major challenges of modern physics~\cite{Bertone:2010zza}. An attractive possibility is to postulate the existence of a dark sector, neutral under the Standard Model (SM) gauge group and interacting with ordinary matter through new mediators or \textit{portals}. Dark photons (or light $Z^\prime$-bosons) induced by hidden gauge groups~\cite{Holdom:1985ag,Fabbrichesi:2020wbt}, or axions and axion-like-particles (ALPs) arising from a spontaneously broken global symmetry  ~\cite{Peccei:1977hh,Peccei:1977ur,Wilczek:1977pj,Weinberg:1977ma,Arvanitaki:2009fg,Kim:2008hd,Jaeckel:2010ni,DiLuzio:2020wdo} are prime examples of bosonic portals (see~\cite{Lanfranchi:2020crw} for a review). Indeed, dark-sector scenarios have attracted much attention over the past years, leading to an extensive experimental program to search for feebly-interacting particles~\cite{Alexander:2016aln,Battaglieri:2017aum,Alimena:2019zri,Beacham:2019nyx}.

If the dark photon is strictly massless, then it can interact with the SM fields only through higher-dimension operators whose structure ultimately depend on the ultraviolet (UV) completion of the model~\cite{Dobrescu:2004wz}. It can couple to fermions of all generations and, in general, mediates flavor-changing processes~\cite{Dobrescu:2004wz,Kamenik:2011vy,Gabrielli:2016cut,Fabbrichesi:2017vma,Fabbrichesi:2019bmo,Su:2019ipw,Su:2020xwt}  (see~\cite{Fabbrichesi:2020wbt} for a review). Also axions and ALPs can display a rich flavor
structure depending, again, on the UV dynamics of the model and, in particular, in scenarios addressing also the flavor puzzle~\cite{Wilczek:1982rv,Feng:1997tn,Calibbi:2016hwq,Ema:2016ops,Bjorkeroth:2018dzu,MartinCamalich:2020dfe,Calibbi:2020jvd}. Hence, rare meson and lepton decays or meson-mixing can pose serious constraints in these models~\cite{Fabbrichesi:2020wbt,MartinCamalich:2020dfe,Calibbi:2016hwq,Calibbi:2020jvd}. 
On the other hand, 
energy-loss arguments applied to stellar evolution lead to some of the strongest indirect bounds on dark sectors~\cite{Sato:1975vy,Vysotsky:1978dc,Dicus:1978fp,Dicus:1979ch,Raffelt:1996wa}. They typically constrain the emission of particles that couple to photons, electrons or nucleons,
and with masses below the temperature of the stellar plasma~\cite{Raffelt:1996wa,Zyla:2020zbs}. A particularly interesting system is the proto-neutron star (PNS) forming during core-collapse supernovae (SN)~\cite{Janka:2006fh}, which reaches temperatures and densities that enable the production of muons~\cite{Bollig:2017lki} or $\Lambda$ hyperons~\cite{Oertel:2016bki}. This opens up the possibility to probe the couplings of the dark sector to heavier flavors of the SM~\cite{Bollig:2020xdr,Croon:2020lrf,MartinCamalich:2020dfe}.

The observation of SN 1987A (and possibly of NS 1987A~\cite{Cigan:2019shp,Page:2020gsx}) has helped to confirm the standard picture of core-collapse SN~\cite{Burrows:1986me,Bethe:1990mw,Janka:2012wk} (see however Ref.~\cite{Bar:2019ifz} for a critical view). An experimental limit on \textit{dark luminosity} stems from the observation of a neutrino pulse, sustained over $\sim 10\,$s~\cite{Loredo:2001rx,Vissani:2014doa}, in coincidence with SN 1987A~\cite{Bionta:1987qt,Hirata:1987hu,Alekseev:1988gp}. Exotic cooling would shorten the neutrino signal, leading to the classical bound~\cite{Raffelt:1996wa},
\begin{align}
\label{eq:Raffelts_Criterium}
L_{\rm d}\lesssim3\times10^{52}~~\text{erg s$^{-1}$},    
\end{align}
at $\sim1 \,$s after bounce (see also~\cite{Raffelt:1987yt,Turner:1987by,Mayle:1987as,Burrows:1988ah,Burrows:1990pk,Hanhart:2000ae,Dent:2012mx,Rrapaj:2015wgs,Chang:2018rso,Carenza:2019pxu,Carenza:2020cis}). 

In the present letter we discuss the possibility that also hyperons can contribute to the dark luminosity through the decay process $\Lambda\to n X^0$ if the dark particles, $X^0$, interact with strange quarks. 
This idea was first explored in~\cite{MartinCamalich:2020dfe} for the case 
that $X^0$ is a flavor-violating QCD axion.
Here we investigate this novel SN cooling mechanism by implementing state-of-the-art simulations combined with proper extensions of the nuclear equations of state (EoS) to include $\Lambda$'s. This allows us to set an upper limit on the branching fraction of the decay that, for definiteness (and simplicity), is applied to the case in which $X^0$ is a massless dark photon or an axion. As we will discuss below, the new SN bound on these models can be orders of magnitude stronger than those obtained from other sources.

\textbf{\textit{Emission rates:}} The width of the decay $\Lambda\to n X^0$ for a massless $X^0$, in vacuum and the $\Lambda$'s rest frame reads
\begin{align}
\label{eq:DecayVacuum}    
\Gamma\equiv\Gamma(\Lambda\to n X^0)=\frac{\bar\omega^3}{2\pi} C_{X},
\end{align}
where $\bar\omega=(m_\Lambda^2-m_n^2)/2m_\Lambda$ is the $X^0$ energy in this frame, $m_{\mathfrak B}$ (${\mathfrak B}=n,~\Lambda$) are the baryon masses and $C_{X}$ is a constant with dimensions of $E^{-2}$ that is related to the energy scale and couplings of the model. The spectrum of the emission rate per unit volume that is induced by this process in the medium is given by
\begin{align}\label{eq:Nexact}
\frac{d\mathcal N_{\rm em}}{d\omega}=&\frac{m_\Lambda^2\Gamma}{2\pi^2\bar\omega} \int^\infty_{E_0} dE \, f_\Lambda(1-f_n),
\end{align}
where $\omega$ ($E$) is the energy of the $X^0$ ($\Lambda$) 
in the PNS's rest frame. The number densities of the baryons follow the relativistic Fermi distributions, $f_\mathfrak{B}$, 
at a given temperature, $T$, and chemical potential, $\mu$, established by ``$\beta$-equilibrium'', $p e^-\leftrightarrow {\mathfrak B}\nu_e$. In Eq.~\eqref{eq:Nexact} we have neglected a Bose-stimulation factor $(1+f_X)$ where $f_X$ is now a Bose-Einstein distribution. Finally, $E_0=m_\Lambda(\omega^2+\bar\omega^2)/(2\omega\bar\omega)$ is the minimal energy of the $\Lambda$ required to produce an $X^0$ with energy $\omega$.
By multiplying Eq.~\eqref{eq:Nexact} by $\omega$ one derives the spectrum of the energy-loss rate $dQ/d\omega$, which integrated over $\omega$ gives the total rate of energy radiated by the star per unit volume of the stellar plasma. 

An approximate (and more intuitive)
formula can be obtained by neglecting the Pauli-blocking for neutrons and taking the limit where $E$, $\omega$, $m_\Lambda - m_n$ are all much smaller than  $m_n$ in Eq.~\eqref{eq:Nexact}~\cite{MartinCamalich:2020dfe},
\begin{align}
\label{eq:Qapp}
Q\simeq n_\Lambda (m_\Lambda-m_n)\Gamma,  
\end{align}
where $n_\Lambda$ is the number density of $\Lambda$ in the medium. If we further neglect interactions of the baryons with the medium, so that $\Lambda$'s are only produced via thermal fluctuations at given chemical potential, then 
\begin{align}
\label{eq:thermal_dist}
n_\Lambda\simeq n_n\exp\left(-\frac{m_\Lambda-m_n}{T}\right).
\end{align}

There are other mechanisms that produce $X^0$ from the $\Lambda n$-coupling,  
such as the \textit{bremsstrahlung} process $\Lambda n\to~n n X^0$. As we will see below, production by $\Lambda \to n X^0$ decays always leads to stronger bounds on the $X^0$ couplings than the corresponding process in nucleons, like $nn\to nnX^0$. Since replacing an initial neutron by a hyperon in this process will only lead to further suppression, the additional contribution to the dark luminosity from $\Lambda$-\textit{bremsstrahlung} can be neglected.

\textbf{\textit{Reabsorption and trapping:}} The emitted $X^0$ can get reabsorbed by the stellar medium if their mean-free path is shorter than the size of the PNS~\cite{1967pswh.book.....Z,1967aits.book.....C,2012sse..book.....K,Raffelt:1987yt,Turner:1987by,Burrows:1990pk}. The main absorption mechanism is the inverse of production, $X^0 n\to\Lambda$, and the absorption rate per unit volume $d\mathcal N_{\rm ab}/d\omega$ can be calculated 
similarly to the emission rate. Assuming time-reversal implies that the matrix elements of both processes are equal while thermal equilibrium implies
that $(1+f_X)f_\Lambda (1-f_n)=f_X(1-f_\Lambda)f_n$. Thus,
\begin{align}
\label{eq:detailed_balance}
\frac{d\mathcal N_{\rm ab}}{d\omega}=\frac{d\mathcal N_{\rm em}}{d\omega},    
\end{align}
which is just the detailed balance between emission and absorption~\cite{1967pswh.book.....Z,1967aits.book.....C,2012sse..book.....K}. From Eq.~\eqref{eq:detailed_balance} it is straightforward to calculate the energy-dependent mean-free path $\lambda_\omega$ as
\begin{align}
\label{eq:free_mean_path}    
\lambda_\omega^{-1}=\frac{1}{\frac{dn_X}{d\omega}}\frac{d\mathcal N_{\rm ab}}{d\omega}=\frac{m_\Lambda^2 \Gamma }{\bar\omega\omega^2}
\int^\infty_{E_0} dE(1-f_\Lambda)f_n,
\end{align}
where $n_X$ is the number density of $X^0$ in the medium. The flux of $X^0$ with energy $\omega$ that propagate outwards in the PNS from a point at radius $r$ will experience an exponential damping from absorption described by the optical depth,
\begin{align}
\tau(\omega,r)=\int^\infty_r\lambda_\omega(r')^{-1}dr',    
\end{align}
where the mean-free path depends on the thermodynamical quantities at $r'$. The total dark luminosity of the PNS can be then written as,
\begin{align}
\label{eq:dark_lumi}
L_{\rm d}=\int d^3\vec{r}\int^\infty_0 d\omega \frac{dQ(r)}{d\omega}e^{-\tau(\omega,r)},  
\end{align}
where the energy-loss rate also depends on the radius. 

This equation describes the attenuation of the flux by re-absorption but it does not account completely for the luminosity in the strong $X^0$-coupling limit, where the mean free path becomes much shorter than the radius of the PNS and the $X^0$ undergoes
multiple absorptions and emissions before leaving the star. In this \textit{trapping regime}, emission of $X^0$ is better described by black-body radiation from a surface where the optical depth is, averaged over $\omega$, equal to 2/3~\cite{1967pswh.book.....Z,1967aits.book.....C}. In our case there is a maximum radius of emission, $R_{\rm d}$, at density and temperature ($T_{\rm d}$) such that $\Lambda$'s are not longer produced in the medium. This sets a minimal emission loss-rate in the trapping regime determined by 
\begin{align}
\label{eq:emission_trapping}
L_{\rm d}^{\rm t}= \frac{\pi^3}{30} g_s R_{\rm d}^2 T_{\rm d}^4,
\end{align}
where $g_s$ the spin-degeneracy factor of the $X^0$.

\textbf{\textit{Supernova simulations and EoS:}} A robust computation of the dark luminosity with the equations above requires knowing the radial profiles of the relevant thermodynamical quantities at a given time of the SN explosion.
We use recent simulations including muons that were developed specifically to constrain the axion-muon coupling using the neutrino data from SN 1987A~\cite{Bollig:2020xdr}. Two EoS are employed for nuclear matter, SFHo~\cite{Steiner:2012rk} and LS220~\cite{Lattimer:1991nc}, and the simulations are performed for different masses of the neutron star spanning the range allowed by observations~\footnote{``SFH'' stands for Steiner, Hempel and Fischer while ``o'' refers to their optimal model~\cite{Steiner:2012rk}. ``LS'' stands for Lattimer and Swesty and the number refers to the nuclear incompressibility $K=220$ MeV that is used~\cite{Lattimer:1991nc}. See   ~\cite{Oertel:2016bki} for a review}. These are labelled by SFHo-18.8, SFHo-18.6 and SFHo-20.0, or by LS220-20.0, depending on the EoS and mass of the progenitor star (in solar masses) used. The simulations are spherically symmetric (one-dimensional) and explosions are, therefore, artificially triggered~\cite{1997PhDT........18R,Rampp:2002bq,Janka:2012wk,Mirizzi:2015eza,Bollig:2020xdr}. The data consist of radial profiles of different thermodynamical variables such as density, temperature and the particle abundancies at various post-bounce times ~\cite{Garching}.         

Hyperons are not included as particle ingredients in the simulations. However, they can be added indirectly through the nuclear EoS because the SFHo and LS220 models have been extended with $\Lambda$'s as explicit degrees of freedom; these EoS are called SFHoY~\cite{Fortin:2017dsj} and LS220$\Lambda$~\cite{Oertel:2012qd,Gulminelli:2013qr}, respectively. In case of SFHo, the hyperonic and non-hyperonic EoS lead to almost identical predictions of the system's thermodynamical properties for all the conditions reached in the SN simulations~\cite{Fortin:2017dsj} (they are also consistent with all known nuclear and astrophysical constraints~\cite{Oertel:2016bki}). Thus, one expects that their results (radial profiles) are not affected by the inclusion of hyperons. In case of LS220, on the other hand, differences between the two types of EoS start occurring at twice nuclear density~\cite{Oertel:2016xsn}, which are conditions reached at the core of the PNS. The LS220$\Lambda$ EoS is also unable to produce neutron stars with $2 M_\odot$ masses, being in conflict with observations~\cite{Demorest:2010bx,Antoniadis:2013pzd,Cromartie:2019kug}. Therefore, in our analysis, we use the results from the SFHo simulations as our baseline and include LS220 only to test the robustness of the results with respect to the choice of EoS.

\begin{figure}[t]
\includegraphics[width=0.48\textwidth]{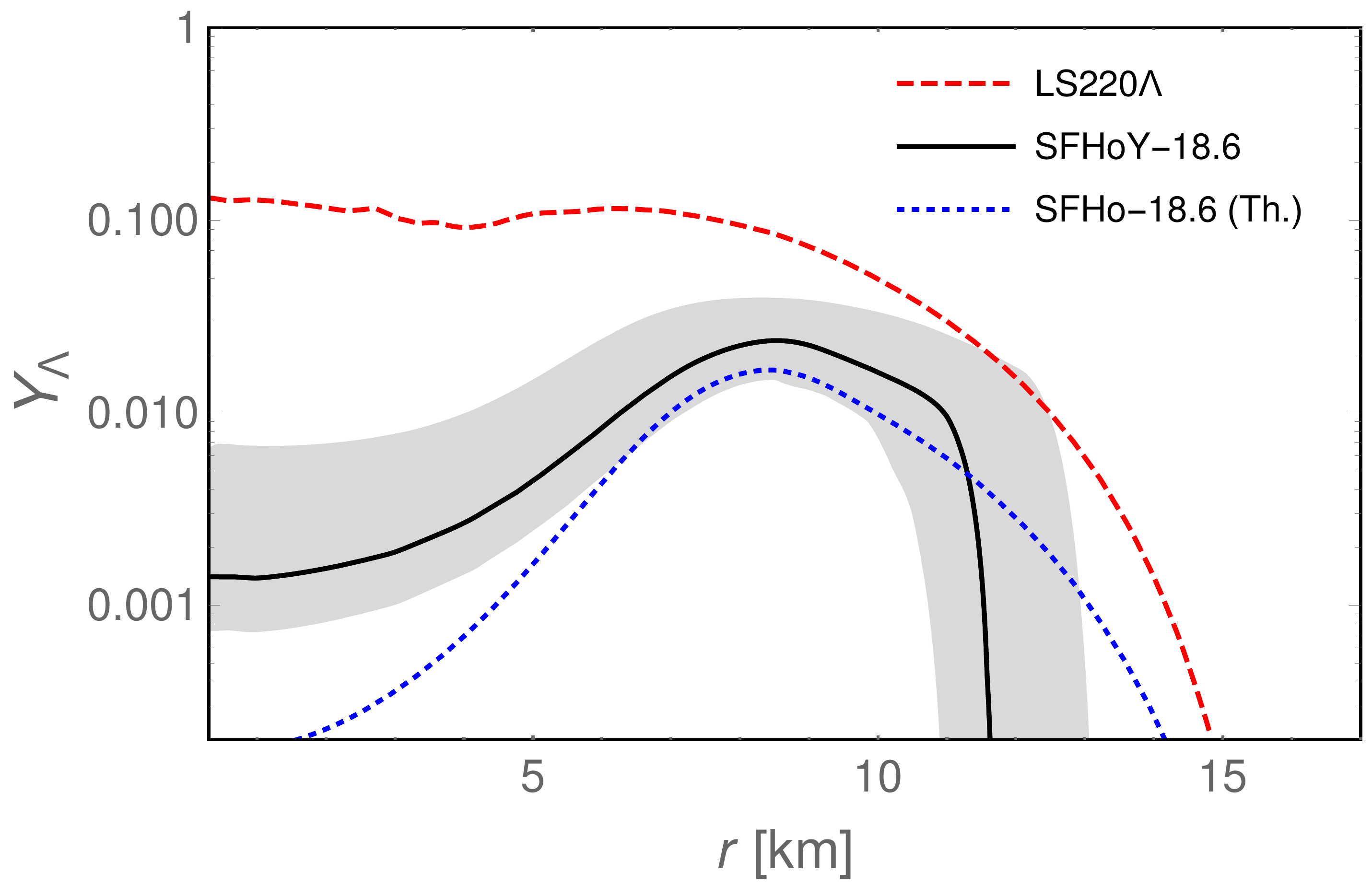} 
\caption{{\small Radial profiles of the $\Lambda$ abundancy $Y_\Lambda = n_\Lambda/n_B$, with $n_B$ the baryon number density,  for the simulations of \cite{Bollig:2020xdr} at $\sim 1\,$s after bounce and using the appropriate extensions of the nuclear LS220 (red dashed) and SFHo EoS with hyperons~\cite{compose}. The upper and lower limits of the gray band around the SFHoY-18.6 curve (solid black) correspond to the SFHo simulations with 20$M_\odot$ and 18.8$M_\odot$, respectively. We also include the results for SFHo-18.6 assuming a purely thermal distribution (blue dotted), see Eq.~\eqref{eq:thermal_dist}.  
\label{fig:Y_Lambda}}} 
\end{figure} 

We use the radial profiles of density, temperature and proton fraction from the simulations as inputs to obtain the radial profiles of the other relevant thermodynamical quantities, which are derived from interpolation tables generated by the CompOSE database~\cite{compose}. In Fig.~\ref{fig:Y_Lambda} we show the profiles of the $\Lambda$ abundancy, $Y_\Lambda$, predicted by these EoS for the simulations of ~\cite{Bollig:2020xdr} at $\sim1 \,$s post-bounce. A large abundancy of $\Lambda$, of the order of 10\% is obtained at the core of the PNS for LS220$\Lambda$. 
In case of SFHoY, $\Lambda$ abundancies are always modest, of less than a few percent, with a maximum at $r\simeq7-8$ km where the PNS reaches the highest temperatures and $\Lambda$-production is dominated by thermal effects.

\textbf{\textit{Medium effects:}} 
SFHoY implements a relativistic microscopic model with baryon-baryon interactions mediated by meson fields that are described in a mean-field approximation. 
For a baryon $\mathfrak B$ with three-momentum $\vec p$, the medium corrections lead to effective masses, $ m_\mathfrak B^*$, and energies, $E_{\mathfrak B}^*=\sqrt{\vec p^{\,2}+m_\mathfrak B^{*2}}+V_{\mathfrak B}$ , where $V_{\mathfrak B}$ is the time-like component of the vector self-energy~\cite{Walecka:1974qa}. 
In the case of LS, the baryon-baryon interactions are modelled using non-relativistic effective interactions. The in-medium masses are not modified while the energies receive a contribution similar to $V_{\mathfrak B}$ but adopting the form of a non-relativistic potential~\cite{Lattimer:1991nc,Oertel:2012qd}.

These medium modifications have to be taken into account in the distributions $f_{\mathfrak B}$ in order to obtain the right baryon abundancies~\cite{compose}, and in the calculation of the emission and absorption rates. These simplify considerably if we neglect $V_{ n}-V_{\Lambda}$, which is $\sim 10$ MeV for all relevant conditions. 
In the Appendix we present the corresponding formulas for $dQ/d\omega$ and $\lambda_\omega^{-1}$.

\begin{table}[t]
\begin{tabular}{ccccc}
\hline\hline\\[-3mm]
 & SFHo-18.6 &  \textbf{SFHo-18.8} &  SFHo-20.0 & LS220-20.0\\
\hline\\[-3mm]
Thermal & $27$& $60$ & $7$ & $6$\\ 
EoS-App. & $20$ & $46$ & $6$ &  $2$\\
EoS  &$36$ & $92$ & $10$ & $4$\\
\textbf{EoS*} & $32$ & $\mathbf{81}$& $9$& $4$\\
\hline\\[-3mm]
$L_{\rm d}^{\rm t}$ [erg s$^{-1}$] &$1.1\times 10^{55}$ & $6.5\times 10^{54}$& $1.7\times 10^{55}$& $1.7\times 10^{54}$\\
\hline\hline
\end{tabular}
\caption{Upper limits on ${\rm BR}(\Lambda\to n X^0)$, in units of $10^{-10}$, for different SN simulations and approaches in the calculation of the dark emissivity (see main text). The value in boldface corresponds to our baseline result. In the last row we show the minimal emissivity achieved in the trapping regime (for $g_s=1$).\label{tab:BR_results}}
\end{table}

\textbf{\textit{Results:}} Combining all the previous ingredients we compute the dark luminosity of SN 1987A as a function of $\Gamma(\Lambda\to n X^0)$. Comparing this to the bound in luminosity shown in  Eq.~\eqref{eq:Raffelts_Criterium} allows us to set an upper limit on the branching fraction of the decay $\Lambda\to n X^0$. In Tab.~\ref{tab:BR_results} we collect our results for the various SN simulations (evaluated at 1 s after bounce) and different approaches to calculate the rates. ``Thermal'' and ``EoS-App.'' employ the approximate Eq.~\eqref{eq:Qapp} in combination with either Eq.~\eqref{eq:thermal_dist} or the corresponding hyperonic EoS for $n_\Lambda$, respectively. ``EoS'' 
is obtained from exactly solving Eq.~\eqref{eq:Nexact} and including medium effects ($m_{\mathfrak B}^*$ and $V_\mathfrak{B}$) in the calculation of the $f_{\mathfrak B}$. In ``EoS*" we also include these effects in the calculation of the rates. 

We find that the bounds are quite robust  with respect to the approach used for the calculation of the luminosities. The largest difference we find is by a factor $\sim2$ in some simulations. On the other hand, they are very sensitive to the mass of the neutron star, with differences that can be larger than by an order of magnitude. The simulations collapsing 20$M_\odot$ provide markedly stronger limits as they are the ones where the PNS reach the highest temperatures and densities~\cite{Bollig:2020xdr}. Although LS220$\Lambda$ predicts a larger number of $\Lambda$'s it eventually leads to only slightly stronger limits compared to SFHoY-20.0.

\begin{figure}[t]
\includegraphics[width=0.48\textwidth]{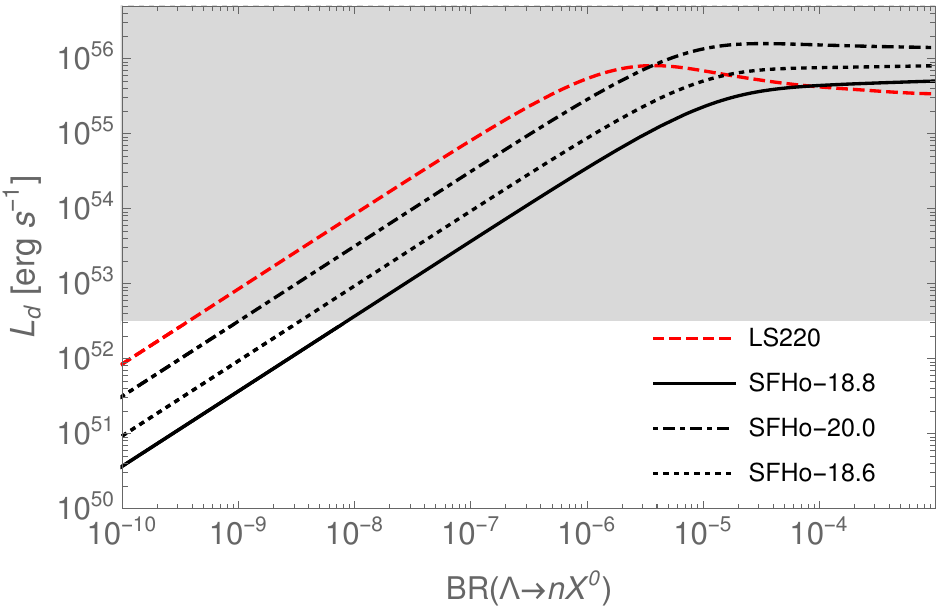} 
\caption{{\small Dark luminosity with EoS* for the various simulations at $\sim1$ s post-bounce as a function of the branching-fraction of the decay $\Lambda\to n X^0$. Gray region is excluded by Eq.~\eqref{eq:Raffelts_Criterium}.
\label{fig:Results}}} 
\end{figure}

In the last row of Tab.~\ref{tab:BR_results} we also show the minimal luminosities obtained in the trapping regime, see Eq.~\eqref{eq:emission_trapping}, which are all much larger than the upper limit in  Eq.~\eqref{eq:Raffelts_Criterium}. This is due to the fact that the last surface of the PNS where $\Lambda$'s can be produced in equilibrium corresponds to a very hot region.
On the other hand, the $X^0$ can still be trapped in the PNS if one adds sufficiently strong interactions with nucleons (star cooling bounds actually prevent trapping with electrons~\cite{Calibbi:2020jvd}). However, the large couplings needed for this are typically excluded by other constraints.  

In Fig.~\ref{fig:Results} we show the dependence of the luminosity on the branching fraction for the different simulations and including medium corrections to the rates. The flattening of the curves at large coupling (or branching ratio) reflects the behavior in the trapping regime discussed previously.
Given all the above, the SN 1987A bound is,
\begin{align}
\label{eq:BRLimit}
{\rm BR}(\Lambda\to n X^0)\lesssim8.0\times 10^{-9},     
\end{align}
obtained by combining the most refined calculation (EoS*) with the simulation giving the weaker bound (SFHo-18.80)~\cite{Bollig:2020xdr}. Note that this is a conservative limit because it stems from the simulation that produces the SN 1987A’s remnant on the low-mass edge of the allowed range and, therefore, have the coolest profile. If we were to use SFHo-20.0 (on the high-mass end of this range) we would get an order-of-magnitude stronger bound, ${\rm BR}(\Lambda\to n X^0)\lesssim9.0\times 10^{-10}$. (See also ref.~\cite{Hanhart:2001fx} for a more refined statistical approach). Finally, let us stress that our constraint is model independent in the sense that it applies to any ultralight dark particle inducing the $\Lambda$ decay and long-lived enough to leave the PNS. 

\textbf{\textit{Dark photons:}} 
In order to apply our result to the massless dark photon case we consider the dimension-five operator
\begin{equation}
\label{eq:LagrangianDP}    
    \mathcal L_{\gamma^\prime}= \dfrac{1}{\Lambda_{\rm UV}}\bar{\psi}_i\sigma^{\mu\nu}\left(\mathbb{C}^{ij}+i \,  \mathbb{C}^{ij}_5\gamma_5\right)\psi_j F^{\prime}_{\mu\nu},
\end{equation} 
where $F^{\prime}_{\mu\nu}$ is the field strength associated to the dark photon, $\psi_i$ are the SM fermions and  $\mathbb{C}^{ij}_{(5)}$
are the couplings of the interaction, suppressed by the energy scale $\Lambda_{\rm UV}$, that depends on the underlying UV completion~\cite{Fabbrichesi:2020wbt}. This operator allows for flavor 
off-diagonal couplings and would contribute to the dark width in Eq.~\eqref{eq:DecayVacuum} with $C_{\gamma^\prime}=8g_T^2/\Lambda_{\rm UV}^2(|\mathbb{C}^{ds}|^2+|\mathbb{C}^{ds}_5|^2)$, where $g_T$ is the $\Lambda\to n$ tensor charge. We use the value $g_T=-0.73$ which is obtained by using SU(3)-flavor symmetry with the tensor charges of the nucleon calculated in the lattice~\cite{Gupta:2018lvp,Georgi:1982jb} (see Appendix for details).

Taking the upper limit on BR$(\Lambda\to n X^0)$ by SN given in Eq. (\ref{eq:BRLimit}) we can set the lower limit,
\begin{align}
\label{eq:LambdaUV}    
\Lambda_{\rm UV}\gtrsim1.2\times10^{10}\text{ GeV},
\end{align}
assuming order-one couplings. 
This can be compared to the limits on flavor-violating hyperon decays from laboratory experiments.
Using the upper bounds on the invisible branching fractions given in Table~II of \cite{MartinCamalich:2020dfe} and the tensor form factors in the Appendix, the decay $\Xi^0\to\Sigma^0\gamma^\prime$ sets the strongest limit, $\Lambda_{\rm UV}\gtrsim4.3\times 10^{7}$ GeV. This could be improved in future experiments like BESIII. Using the prospected sensitivity for BR$(\Lambda\to~n \gamma^\prime)\simeq {\rm BR}(\Lambda\to n \nu\bar\nu)$~\cite{Li:2016tlt}, the bound could be pushed up to $1.9\times 10^9$ GeV. Kaon decays can place a similar limit using  BR$(K^+\to \pi^+\pi^0X^0)$ \cite{Adler:2000ic}; applying the calculations derived in \cite{Su:2020xwt} we get $\Lambda_{\rm UV}\gtrsim1.7\times 10^{7}$ GeV. Probing this beyond the SN limit, for example at the NA62 experiment~\cite{Fabbrichesi:2017vma}, would require reaching a sensitivity  BR$(K^+\to\pi^+\pi^0 X^0)\lesssim1.85\times10^{-10}$.

\begin{figure}[t]
\includegraphics[width=0.48\textwidth]{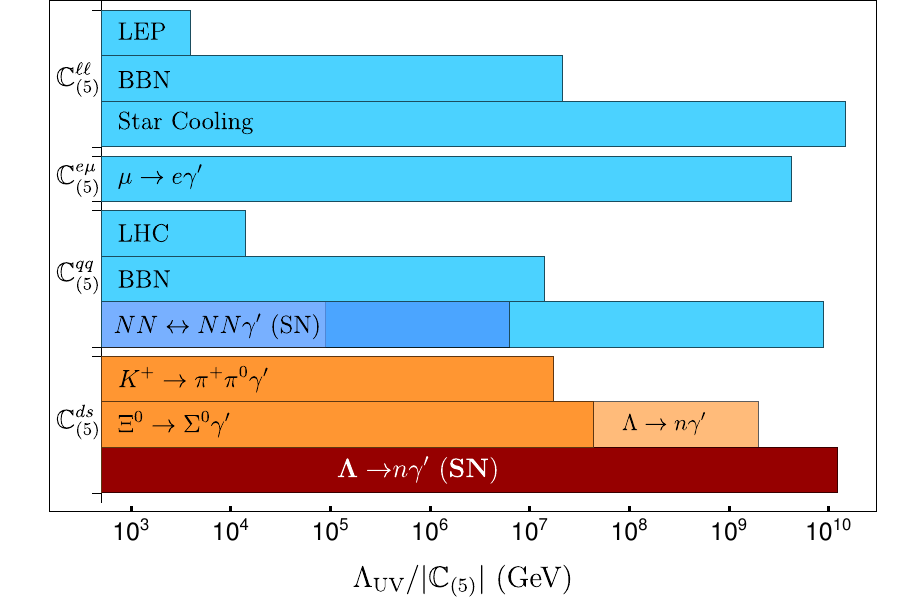} 
\caption{{\small Model independent excluded region
of the couplings of the dipole operator Eq.~(\ref{eq:LagrangianDP}) from various sources. In blue, limits to quark flavor diagonal and lepton couplings; in orange, limits to $ds$ coupling that can be directly compared to the one derived in this work, in dark red. In faint orange the prospected limit from BESIII.
\label{fig:DPlimits}}} 
\end{figure}

We can also compare with bounds on dark photon couplings to other matter fields, which have been collected in  Ref.~\cite{Fabbrichesi:2020wbt}. In Fig.~\ref{fig:DPlimits} we show these limits together with the new SN bound on $ds$-couplings (note that we have updated the SN bounds on nucleons using Ref.~\cite{Carenza:2019pxu}). We see that the SN analysis done in this work sets the strongest limit on dark photon couplings, along with star cooling constraints on lepton couplings. This bound cannot be avoided by trapping the dark photon with large couplings to nucleons, since constraints from SN 1987A close the entire window, see Fig.~\ref{fig:DPlimits}. This results from combining the luminosity constraint (light blue) with the absence of a signal in the Kamiokande detector (dark blue).

\textbf{\textit{Axions:}} The couplings of axions to SM fields are
\begin{align}
\label{eq:axion_couplings}
& \mathcal{L}_{a} = 
\frac{\partial_\mu a}{2 f_a} \, \bar \psi_i \gamma^\mu \big(c^V_{i j}+c^A_{ij} \gamma_5 \big) \psi_j \, , 
\end{align}
where $a$ is the axion field and $f_a$ is its decay constant. The axion contribution to Eq.~\eqref{eq:DecayVacuum} is given by $C_a=(f_1^2|c^V_{ds}|^2+g_1^2|c^A_{ds}|^2)/(2 f_a)^2$, where $f_1$ and $g_1$ are form factors that are discussed in the Appendix. Using the values shown there and the SN limit in Eq.~\eqref{eq:BRLimit} we obtain,
\begin{align}
\label{eq:Faxion}
F^V_{sd} \gtrsim7.1\times 10^9\text{ GeV,}~~~~F^A_{sd} \gtrsim5.2\times 10^9\text{ GeV},
\end{align}
for pure vector and axial couplings $F^{V,A}_{sd} \equiv 2 f_a/c^{V,A}_{sd}$, respectively. A comprehensive discussion of other bounds in this model can be found in~\cite{MartinCamalich:2020dfe}. In particular, this constraint is also stronger than the SN bounds on the diagonal couplings to light quarks from nucleon-nucleon \textit{bremsstrahlung},  and on the leptonic couplings to $\mu\mu$ \cite{Bollig:2020xdr} and $\mu e$ \cite{Calibbi:2020jvd}.  Finally, as we discuss in detail in the Appendix, the bounds for the QCD axion are roughly applicable also to the ALP case 
unless its mass is very close to the $\Lambda- n$ mass difference. It is interesting to note that 
the SN bound on ALPs can become comparable to the 
stringent bounds from laboratory experiments looking for $K^+ \to \pi^+ X^0$ in the 
two-pion decay region, where the sensitivity is strongly reduced due to the SM background~\cite{CortinaGil:2020zwa}.

\textbf{\textit{Conclusions:}} We have studied in detail a novel SN bound on dark flavored sectors stemming from the decay of $\Lambda$ hyperons in the proto-neutron star. We have used state-of-the-art simulations with the corresponding hyperonic EoS for our calculations to obtain the upper limit BR$(\Lambda\to nX^0)\lesssim8.0\times10^{-9}$. This leads to  the strongest
bounds that have been derived so far on the couplings of the massless dark photon to quarks. 
This analysis also sets strong constraints 
on flavor-violating axion models, and 
can be readily extended to other flavored dark sectors.


\textbf{\textit{Acknowledgements:}} We thank R.~Bollig, H.-Th.~Janka and M.~Oertel for useful discussions and for providing us with internal details of their work. We also thank F.~S.~Kitaura for useful discussions. JMC acknowledges support from the Spanish MINECO through the ``Ram\'on y Cajal'' program RYC-2016-20672 and the grant PGC2018-102016-A-I00. The work of JTC is supported by the Ministerio de Ciencia e Innovaci\'on under FPI contract PRE2019-089992 of the SEV-2015-0548 grant and Generalitat Valenciana by the SEJI/2018/033 project. L.T. acknowledges support from the Ministerio de
Econom\'ia y Competitividad under contract FPA2016-81114-P, the Ministerio de Ciencia e Innovaci\'on under contract FIS2017-84038-C2-1-P, by PHAROS COST Action CA16214, and by the EU STRONG-2020 project under the program  H2020-INFRAIA-2018-1, grant agreement no. 824093. This work is partially supported by project C3b of the DFG-funded Collaborative Research Center TRR257,  ``Particle  Physics  Phenomenology  after  the  Higgs  Discovery".

\vspace{0.4cm}
\textbf{\textit{\Large Appendix}}
\vspace{0.4cm}

\textbf{\textit{Baryon form factors:}} The baryonic matrix elements required for the $\mathfrak B_1\to \mathfrak B_2 X^0$ transition are 
\begin{align}
\label{eq:baryons_FFs}
\langle \mathfrak B_2(p^\prime) | &\bar{d} \sigma^{\mu\nu} s |\mathfrak B_1 (p)\rangle 
=g_T\bar u_2(p^\prime)\sigma^{\mu\nu}u_1(p),\nonumber\\
\langle \mathfrak B_2(p^\prime) | &\bar{d} \gamma_\mu s |\mathfrak B_1 (p)\rangle
=
\bar{u}_2 (p^\prime)  \Big[
f_1(q^2)  \,  \gamma_\mu  
\nonumber\\  
&+ \frac{f_2(q^2)}{m_{\mathfrak B_1}}   \, \sigma_{\mu \nu}   q^\nu  
+ \frac{f_3(q^2)}{m_{\mathfrak B_1}}   \,  q_\mu  
\Big]  
 u_1 (p),\nonumber \\ 
\langle \mathfrak B_2(p^\prime) | &\bar{d} \gamma_\mu \gamma_5 s |\mathfrak B_1 (p)\rangle
=
\bar{u}_2(p^\prime)  \Big[
g_1(q^2)    \gamma_\mu
\nonumber\\    
&+ \frac{g_{2} (q^2)}{m_{\mathfrak B_1}}   \sigma_{\mu \nu}   q^\nu  
+ 
\frac{g_{3} (q^2)}{m_{\mathfrak B_1}}   q_\mu  
\Big]  \,\gamma_5  u_1(p),
\end{align}
where $q=p-p'$. The constant $g_T$ is the tensor charge. It enters in the amplitude of the decays to the massless dark photon. A matrix element of the tensor operator with a $\gamma_5$ is related to $g_T$ via the relation $2\sigma^{\mu\nu}\gamma_5=i\,\varepsilon^{\mu\nu\alpha\beta}\sigma_{\alpha\beta}$. The functions $f_i(q^2)$ and $g_i(q^2)$ are the vector and axial-vector form factors that enter in the decays to axions and which depend on $q^2=m_a^2$. Note that only $f_1(q^2)$, $f_3(q^2)$, $g_1(q^2)$ and $g_3(q^2)$ contribute after contracting by $i q^\mu$. Furthermore, only the charges $f_1\equiv f_1(0)$ and $g_1\equiv g_1(0)$ are needed for the massless axion. 

In the SU(3)-flavor symmetric limit of QCD there are only two independent reduced matrix elements for each current (or Lorentz structure)~\cite{Georgi:1982jb}, which can be determined by using inputs from experiment or lattice QCD. In case of the vector and axial-vector charges one uses the baryon's electromagnetic charges and form factors measured in semileptonic hyperon decays, respectively  (see~\cite{MartinCamalich:2020dfe}). For the tensor charges, one uses lattice QCD calculations as input~\cite{Gupta:2018lvp}. In Tab.~\ref{tab:gT} we show the tensor charges for different baryon transitions triggered by the couplings $\mathbb{C}_{(5)}^{ds}$ in Eq.~\eqref{eq:LagrangianDP}.


\begin{table}[h]
  \begin{tabularx}{.48\textwidth}{ >{\setlength\hsize{1\hsize}\centering}X>{\setlength\hsize{1\hsize}\centering}X>{\setlength\hsize{1\hsize}\centering}X>{\setlength\hsize{1\hsize}\centering}X>{\setlength\hsize{1\hsize}\centering}X>{\setlength\hsize{1\hsize}}X } 
    
    \hline\hline

    $\Lambda n$& $\Sigma^+ p$ & $\Sigma^0 n$ & $\Xi^-\Sigma^-$ & $\Xi^0\Sigma^0$ & \quad$\Xi^0\Lambda$ \\
    \hline\\[-3mm]
    $-0.73$ & $0.20$ & $-0.14$ & $0.99$ & $-0.70$ & \quad$0.24$\\
    \hline\hline
  \end{tabularx}
\caption{Baryon tensor charges obtained using SU(3)-flavor symmetry and lattice QCD results~\cite{Gupta:2018lvp}.\label{tab:gT}}
\end{table}
Finally it is important to stress that these predictions of the matrix elements are accurate up to SU(3)-breaking effects~\cite{MartinCamalich:2020dfe}, which are suppressed by  $\sim(m_\Lambda-m_n)/m_ \Lambda\simeq0.15$.

\textit{\textbf{Supernova bounds for ALPs:}} The calculation of the energy-loss rate by ALP emission is analogous to the one for the axion. The width in Eq.~\eqref{eq:DecayVacuum}, for the decay $\Lambda\to n a$ in vacuum and $\Lambda$'s rest frame, is generalized,
\begin{align}
\label{eq:width_ALP}
\Gamma_a=\frac{\bar p\,\bar\omega^2}{2\pi} C_a, 
\end{align}
where $\bar p=\lambda^{1/2}(m_\Lambda^2,m_n^2,m_a^2)/2m_\Lambda$ is now the modulus of the 3-momenta in the decay, with $\lambda(x,y,z)$ the K\"all\'en function. The constant $C_a$ is
\begin{align}
\label{eq:Ca}
C_a=\left(1-\frac{x_a^2}{(2-\delta)^2}\right)|\mathcal V|^2+\left(1-\frac{x_a^2}{\delta^2}\right)|\mathcal A|^2&.
\end{align}
We have defined the dimensionless quantities $\delta=(m_\Lambda-m_n)/m_\Lambda$, $x_a=m_a/m_\Lambda$, and the functions
\begin{align}
\label{eq:V&A}
\mathcal V=&\frac{c^V_{ds}}{2 f_a}\left(f_1(q^2)+\frac{q^2}{m_\Lambda(m_\Lambda-m_n)}f_3(q^2)\right),\nonumber\\    
\mathcal A=&\frac{c^A_{ds}}{2 f_a}\left(g_1(q^2)-\frac{q^2}{m_\Lambda(m_\Lambda+m_n)}g_3(q^2)\right),
\end{align}
where we have used the Lagrangian in Eq.~\eqref{eq:axion_couplings} and the form factors in Eq.~\eqref{eq:baryons_FFs}, and where $q^2=m_a^2$. 

These equations can be simplified by expanding around the SU(3)-flavor symmetric limit.  
We take $\delta$ as the expansion parameter, which controls the phase space of the decay because $q^2/m_\Lambda^2=x_a^2\leq \delta^2$. 
The form factors can be also expanded in $q^2/m_R^2\sim\delta^2$,
where $R$ is a suitable hadronic resonance coupling to the $sd$ current~\cite{Ecker:1989yg,Masjuan:2012sk}. 

\begin{figure}[t]
\includegraphics[width=0.48\textwidth]{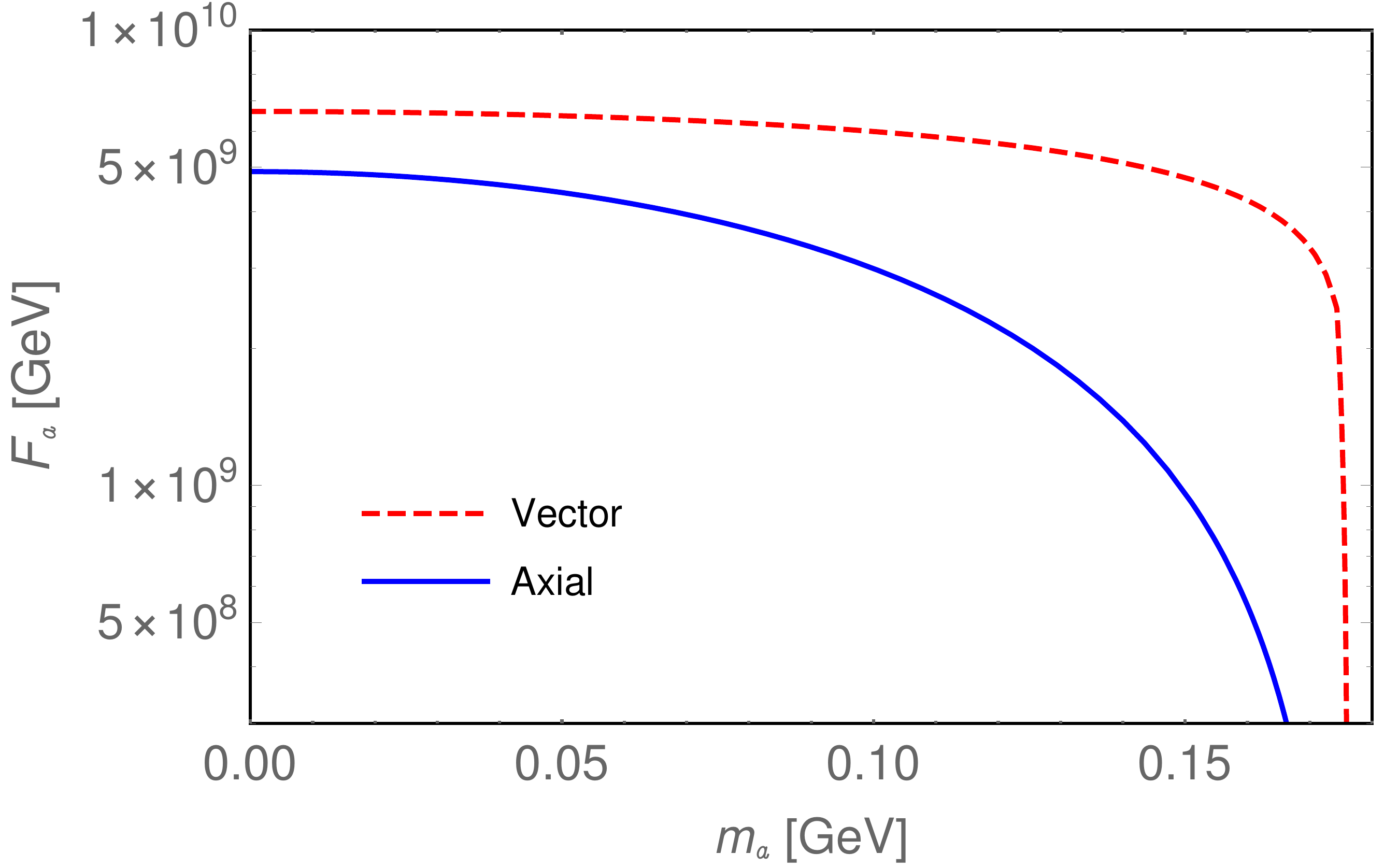} 
\caption{{\small Dependence  on the ALP mass of the lower bounds of $F_a = 2 f_a/c^{V,A}_{sd}$ for the vector or axial couplings (only one active at the time). The calculations have been done in the EoS approach using the SFHo-18.8 simulation.  
\label{fig:ALP}}} 
\end{figure}

Expanding Eq.~\eqref{eq:Ca} one finds the simpler form,
\begin{align}
\label{eq:Ca_app}
C_a=\frac{1}{(2 f_a)^2}\left(|c^V_{ds}|^2f_1^2+|c^A_{ds}|^2(1-\frac{x_a^2}{\delta^2})g_1^2\right)+\mathcal O (\delta^2),
\end{align}
where we have also used the fact that $f_3(q^2)\sim\mathcal O(\delta)$~\cite{Weinberg:1958ut}. Note that in this approximation the only information on the baryonic matrix elements that is needed is the same as in the case of the massless axion. 

We can now turn to the calculation of the volume energy-loss rate. We find it convenient to express it as the double integral in the ALP's and the $\Lambda$'s energies,
\begin{align}
Q=\frac{m_\Lambda^2\Gamma_a}{2\pi^2\bar p}\int_{m_\Lambda}^\infty dE\int_{\omega_-}^{\omega_+}d\omega\omega f_\Lambda(1-f_n),    
\end{align}
where the limits in $\omega$ are 
\begin{align}
\label{eq:omegalimits}
\omega_\pm=\frac{1}{2m_\Lambda^2}\left((m_\Lambda^2+m_a^2-m_n^2)E\pm2m_\Lambda\bar p p\right),
\end{align}
with $p=\sqrt{E^2-m_\Lambda^2}$.

In Fig.~\ref{fig:ALP} we show the dependence of the SN bounds on the mass of the ALP. The calculations have been done in the EoS approach and using the simulation SHFo-18.8. As we can see in the plot, the bounds obtained in the massless case are a good representative of the bounds obtained for the ALPs in most part of the allowed mass range. The axial coupling is more sensitive to the ALP's mass, see Eq.~\eqref{eq:Ca_app}, and the corresponding constraint decreases by more than an order of magnitude when $m_a\gtrsim0.15$ GeV.

\textit{\textbf{Rates including medium corrections}:} 
In the SFHoY EoS, the difference between the time-like components of the vector self-energy for $\Lambda$ and $n$ are within $\sim 10$ MeV for all the conditions considered in this work~\cite{compose}. We keep this difference for the calculation of the particle distributions of the baryons (to obtain the right baryon abundancies), but we use the approximation $V\equiv V_n\approx V_\Lambda$ in the calculation of the rates. The spectrum of the volume emission is expressed as an integral in the $\Lambda$'s energy ($E$),
\begin{align}\label{eq:Q_medium}
\frac{dQ}{d\omega}=&\frac{\Gamma\,\omega}{8\pi^2\bar \omega^3}\left(m_\Lambda^{*2}-m_n^{*2}\right)^2\nonumber\\
&\times\int^\infty_{E_0^*} dE \frac{(E-V)(E-V-\omega)}{E(E-\omega)}\,f_\Lambda(1-f_n),
\end{align}
where $E_0^*=m_\Lambda^*(\bar\omega^{*2}+\omega^2)/2\bar\omega^*\omega+V$ and $\bar \omega^*=(m_\Lambda^{*2}-m_n^{*2})/2m_\Lambda^*$. The corresponding mean free path is,
\begin{align}
\label{eq:free_mean_path_medium}
\lambda_\omega^{-1}&=\frac{\Gamma }{4\bar\omega^3\omega^2}\left(m_\Lambda^{*2}-m_n^{*2}\right)^2\nonumber\\
&\times\int^\infty_{E_0^*} dE \frac{(E-V)(E-V-\omega)}{E(E-\omega)}\,(1-f_\Lambda)f_n.
\end{align}

In case of LS220, the baryon masses do not change and $V_{\mathfrak B}$ adopts the form of a nonrelativistic potential. The difference between these potentials for $\Lambda$ and neutron is typically of $\sim 10-20$ MeV reaching, at most, $\sim100$ MeV at 3-4 times nuclear saturation density~\cite{compose}. Thus, for LS220 we keep this difference for the baryon distributions and use Eqs.~\eqref{eq:Q_medium}  and~\eqref{eq:free_mean_path_medium}, but replacing $m_{\mathfrak B}^*$ by $m_{\mathfrak B}$ and interpreting $V_{\mathfrak B}$ as the nonrelativistic potentials.

Finally, we neglect medium corrections to the baryon form factors. As discussed earlier, they already carry a significant uncertainty stemming from the SU(3)-flavor symmetry used to derive them.

\bibliographystyle{apsrev4-1}
\bibliography{SNflavored.bib}
\end{document}